\documentclass[]{aa} 

\usepackage{natbib}
\usepackage{graphicx}
\bibpunct{(}{)}{;}{a}{}{,}
\usepackage{txfonts}
\usepackage[dvipsnames,usenames]{color}
\begin{document}

\title{A multi-epoch spectroscopic study  of the BAL quasar\\ APM 08279+5255: I. C {\Large IV} absorption  variability}

\author{Trevese, D.
         \inst{1},
               Saturni, F.G.
         \inst{1},
        Vagnetti, F.
         \inst{1,2},
        Perna, M.
          \inst{1},
        Paris, D.
          \inst{3},
        Turriziani, S.
          \inst{2}
          }

  \institute{Dipartimento di Fisica, Universit\`a di Roma La Sapienza, Piazzale Aldo Moro, 5, I-00185 Roma (Italy)\\
              \email{dario.trevese@roma1.infn.it}
         \and Dipartimento di Fisica, Universit\`a di Roma Tor Vergata, Via della Ricerca Scientifica, 1, I-00133 Roma (Italy) 
         \and Osservatorio Astronomico di Roma, Via Frascati, 33, I-00040 Monte Porzio Catone (Italy)
             }

   \date{}
 
  \abstract
    {Broad Absorption Lines  indicate gas outflows with velocities from thousands km s$^{-1}$ to about 0.2 the speed of light, which may be present in all quasars and may play a major role in the evolution of the host galaxy. The variability of absorption patterns can provide informations on changes of the density and velocity distributions of the absorbing gas and its ionization status. }
  {We want to follow accurately the evolution in time of the luminosity and both the broad and narrow C {\scriptsize IV} absorption features  of an individual object, the quasar APM 08279+5255, and analyse the correlations among these quantities. }
   {We collected 23 photometrical and spectro-photometrical observations at the 1.82 m Telescope of the Asiago Observatory since 2003, plus other 5 spectra from the literature. We analysed the
   evolution in time of the equivalent width of the broad absorption feature and two narrow absorption systems, the correlation among them and with the R band magnitude. We performed a structure function analysis of the equivalent width variations. }
   {We present an unprecedented monitoring  of a broad absorption line quasar based on 28 epochs in 14 years. The shape of  broad absorption feature shows a relative stability, while its equivalent width slowly declines until it sharply increases during 2011. In the same time the R magnitude stays almost constant until  it sharply increases during 2011. The equivalent width of the narrow absorption redwards of the systemic redshift only shows a decline.}
  {The broad absorption behaviour suggests changes of the ionisation status as the main cause of variability. We show for the first time a correlation of this variability with the R band flux. The different behaviour of the narrow absorption system might be due to recombination time delay. The structure function of the absorption variability has a slope comparable with typical optical variability of quasars. This is consistent with variations of the  200~\AA~ ionising flux originating in the inner part of the accretion disk.}

\authorrunning{D. Trevese et al.} 
\titlerunning{C {\scriptsize IV}  absorption variability in  APM 08279+5255}
   \maketitle

\section{Introduction}
BAL QSOs were discovered by \citet{Lynd67} and their properties are described  in \citet{Weym83,Turn88}.
The spectra of about 20\% of all quasars (QSOs) exhibit broad absorption lines (BALs)  with velocities from thousands km s$^{-1}$ to $\la$ 0.2  the speed of light, indicating the outflow of material and energy from the active galactic nucleus (AGN) to the surrounding space \citep{Hewe03,Gibs08,Cape11}. 
Mechanical and radiative energy transfer to the QSO environment can affect the galaxy evolution processes,   so that  understanding gas outflows has become crucial in establishing the role played by the AGN feedback  in the cosmological process of galaxy formation and evolution \citep[][and refs. therein]{Catt09}, as suggested, in particular, in the case of the "ultra-fast outflows" detected in X-rays  in  several $z<0.1$ AGNs
\citep{Tomb10}
 and specifically in APM 08279+5255 
\citep{Char09}, 
which is the subject of the present study.
The fact that BALs appear  only in a fraction of QSO spectra, may indicate that either they are seen only in particular directions with respect to the axis of the accretion disk \citep[and refs. therein]{Elvi00}, or they are observed during particular phases of the QSO life \citep[][and refs. therein]{Farr07}.
Investigating the nature and location of these outflowing absorbers could, in principle, provide clues
to understand the dynamics and ionisation processes in the circum-nuclear gas.   
Unfortunately, structure,  location, dynamics and ionisation state of the absorbers are poorly known so far.
Variability of BALs can provide further important informations about the properties of the absorbing gas. 
BAL variations were first detected by \citet{Folt87} \citep[see also][and refs. therein]{Barl92}.
In principle, variability may be caused by the motion of  gas clouds across the line of sight,  or by changes of the ionisation status of the gas.  A first systematic analysis of BAL variability was presented by  \citet{Barl93}, who analysed 23 BALQSOs, each observed typically 2--3  times in the course of four years. Continuum variations were associated with BAL variation only in some cases, suggesting the possibility that changes in the ionisation status could cause absorption variability, possibly with some phase difference between  the far UV ionising flux  and the observed continuum changes.
 Since then, some other systematic studies were devoted to the analysis of BAL variability.
They are mainly focused on C {\scriptsize IV} BAL which  is  usually the most visible absorption feature.
\citet{Lund07} used the Sloan Digital Sky Survey (SDSS; \cite{York00}) spectra of 29 BAL QSOs,  observed twice in a total observing period about one year,  corresponding to rest-frame time lags in the range two weeks- four months. Their analysis is limited to those BALs which are  separated from the emission peak by more than 3600 km s$^{-1}$, to avoid the complication due to the emission line variability. They observe that the strongest BAL variability occurs among absorption features with the smallest equivalent widths and with velocities exceeding 12,000 km s$^{-1}$. They conclude that  strong variability at high velocity might be consistent with Kelvin-Helmholtz instabilities predicted in disk-wind models of \citet{Prog00}.
 \citet{Gibs08} studied the BAL variations on long timescales of 13 QSOs, in the redshift range 1.7--2.8, taking advantage of the overlap between the Large Bright Quasar Survey (LBQS; \citep{Hewe95} and refs. therein) and SDSS which observed these objects with time differences of 10--18 years, corresponding to  rest-frame time lags in the range 3--6 years. They found that the equivalent width variations tend to increase with the time lag, in the observed interval. They found no evidence that variations are dominated by changes in the photo-ionisation on multi-year timescales.
A subsequent study of \citet{Gibs10} analysed the variations of 9 BALQSOs
in 2--4 epochs, considering both C {\scriptsize IV} and Si {\scriptsize IV} BALs. They found some correlation between C {\scriptsize IV} and Si {\scriptsize IV} equivalent width variations, estimated lifetimes $\ga$ 30 yr for the stronger BALs, and do not find asymmetries in the growth and decay timescales.
\citet{Cape11} report the first results of  new observations of the \citet{Barl93} sample, obtained with the 2.4 m Hiltner telescope of the MDM Observatory, supplemented with SDSS spectra, for a total of 120 spectra for 24 objects.
This study provides a comparison among short and long time scale variations and indicates a dependence of variability on the velocity of the outflow and the BAL strength.
A subsequent study by \citet{Cape12}, presents further results on their monitoring of 24  BALQSOs and compares C {\scriptsize IV} and Si {\scriptsize IV} BAL variability, suggesting a complex scenario that seems to favour ionisation changes of the outflowing gas while at the same time variations in limited portions of the broad absorption troughs may indicate movements of the individual clouds across the line of sight. 
A further contribution by \citet{Cape13} is focused on short time scale variability of the C {\scriptsize IV} BAL from the same sample. Variations in only portions of the  BAL suggest the presence of substructures in the outflow, moving across the line of sight, and provide constrains on the speed and geometry of the outflows. 
Clearly, the amount of data, both in terms of number of different objects and sampling frequency of the individual quasars, is crucial to assess the properties of BAL variability. Sampling of individual QSOs is, however, limited by the intrinsic timescales of variations, especially for high redshift objects.
In the present study, we discuss the result of the monitoring of a single object, APM 08279+5255 at $z$=3.911, which we observed 23 times since 2003, in the framework of a monitoring campaign of medium-high redshift QSOs, carried out with the 1.82 m  Copernico telescope of the Asiago Observatory and
devoted to determine the mass of the central supermassive black hole (SMBH) through reverberation mapping \citep{Trev07}.  Since APM 08279+5255 is one of the brightest QSOs in the sky, and shows  several other interesting peculiarities, it was observed several times by other authors, photometrically and spectroscopically, soon after its discovery \citep{Irwi98}. So that it was possible to collect both spectroscopic and photometric data from the literature, which have  been included in the present analysis, providing what is, to our knowledge, the best sampling of  the variability of a BAL QSO available to date. In section 2 we describe the data we collected from the literature, in section 3 we describe our spectrophotometric monitoring campaign and data reduction, in section 4 we analyse the variability of the C {\scriptsize IV} absorption systems and its correlation with photometric variability and in section 5 we discuss and summarise the results.

\section{Data from the literature}
\subsection{Spectroscopy}
The BAL QSO APM 08279+5255 was serendipitously discovered in the course of a survey conducted with the 2.5 m Isaac Newton Telescope (INT), La Palma, for the identification of cool carbon stars in the Galactic halo \citep{Irwi98}.
Observations were  taken between February 27 and March 6, 1998 with a resolution $\lambda / \Delta \lambda\sim 2000$, and a redshift $z=3.87$ was initially attributed on the basis of the Si {\scriptsize IV}+O {\scriptsize IV]} feature, the C {\scriptsize III]} $\lambda 1909$~\AA~ at $\sim 9300$~\AA~ and N {\scriptsize V} $\lambda 1240$ at $6040$ \AA. M.J. Irwin kindly provided us with the data of the discovery spectrum, which are included in the present analysis.  This object, with a magnitude R=15.2, appeared as the brightest QSO in the sky, with a bolometric luminosity  exceeding $\sim 5 \times 10^{15} L_{\odot}$, and has been the subject of a great number of papers, of which we mention only those which are relevant for the specific purpose of our study.
Due to its high redshift, APM 08279+5255 is an ideal ``background'' source for analyses of the intervening absorptions. In fact, high resolution (R $\approx$ 50,000) spectra were taken with the HIRES spectrograph at Keck telescope for the analysis of the Ly$\alpha$ clouds  distribution and metal abundances \citep{Elli99,Elli04}.
A low  resolution spectrum   ($\lambda / \Delta \lambda\sim 400$) was obtained  by  \citet{Hine99}, together with polarised spectra which permitted the identification of a broad  O {\scriptsize VI} $\lambda\lambda1032, 1038$~\AA~ absorption, associated with Ly$\beta$, falling in the Ly$\alpha$ forest. Also this spectrum, kindly provided by D.C. Hines, is included in the present analysis.
Soon after the discovery, adaptive optics images at CFHT telescope have shown APM 08279+5255 to be a double source, likely gravitationally lensed \citep{ledo98}, as already proposed by \citet{Irwi98}, and Hubble Space Telescope (HST)  and radio images suggested the existence of  a long sought ``third component'', expected in the case of axi-symmetric gravitational lensing fields  \citep{Ibat99}. A lens model indicating a magnification of about 100 was derived by \citet{Egam00}. Spectroscopic  HST  observations, at medium-high  resolution ($\lambda / \Delta \lambda\sim 4500)$ confirmed that the third component is in fact part of the lensed image \citep{Lewi02}.
Keck and HST spectra, besides providing two more epochs for the spectroscopic monitoring, are also used in our analysis to resolve and identify absorption features not resolved in our spectra.

\begin{figure}
   \centering
\resizebox{\hsize}{!}{\includegraphics{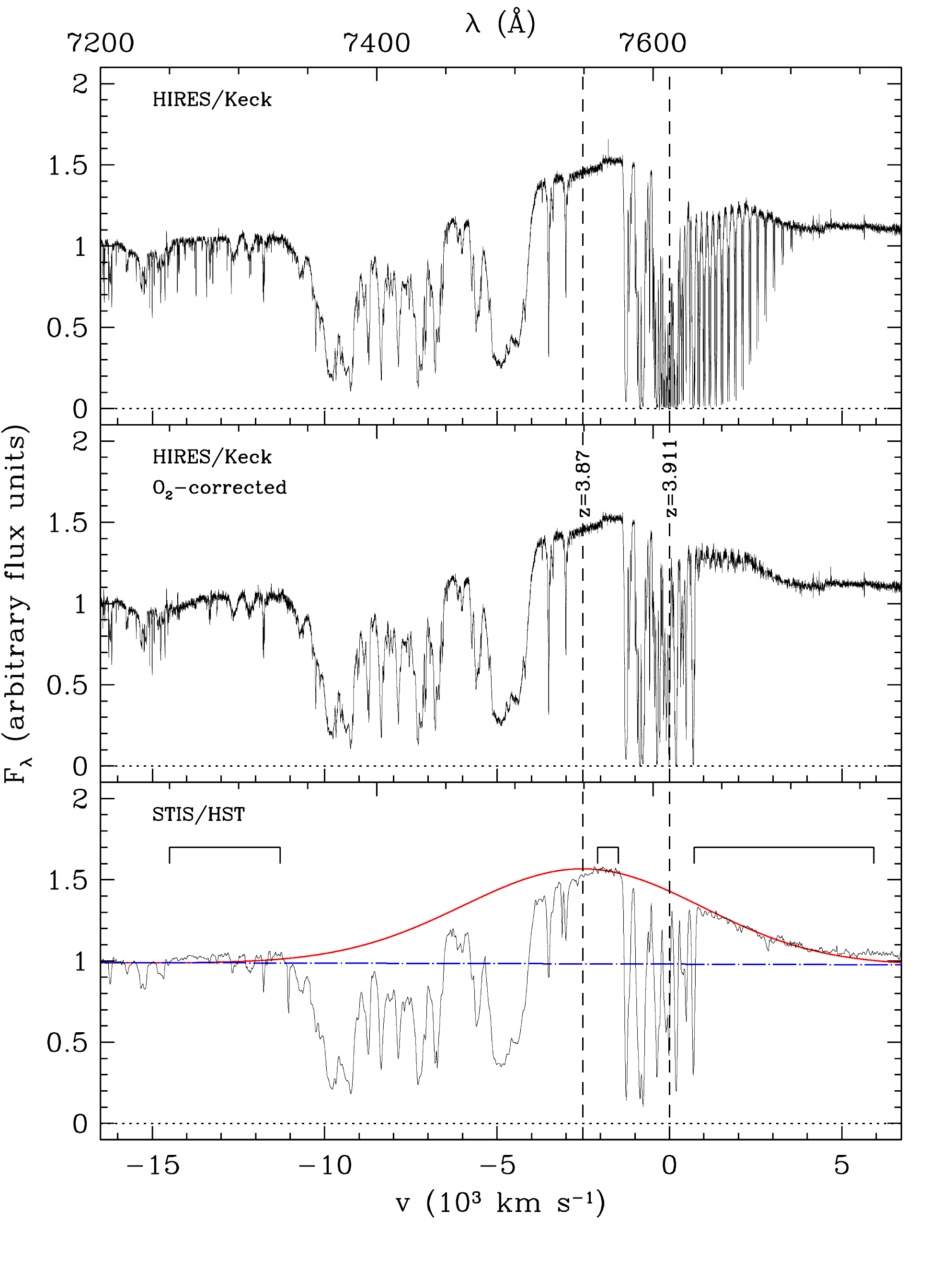}}
      \caption{ 
     The  (uncalibrated) flux in the  $\lambda\lambda 7200-7780$~\AA~region. {\it Top panel}: from the Keck HIRES high resolution spectrum \citep{Elli99}; {\it middle panel}: the same spectrum after correction for telluric $O_2$ band (see Section 3); {\it bottom panel}: from  HST STIS spectrum \citep{Lewi02}, where the the continuum (blue dashed line) and a Gaussian fit to the C IV emission line (red curve) are shown. A comparison of  middle and bottom panels shows that the O$_2$ correction is satisfactory. The velocity scale is based on the systemic redshift $z$=3.911 \citep{Down99}.
}
  \label{fig_1-sp-keck-hst}
   \end{figure}

The HST spectrum has been obtained  by \citet{Lewi02} with STIS.
High resolution spectra allowed us to identify the absorption-free regions, shown in Figure  \ref{fig_1-sp-keck-hst}, which can be used to fit the C {\scriptsize IV} emission line profile. The same figure also shows two narrow absorption line (NAL) systems: the first, identified by \citet{Elli04}, that we will call ``red-NAL'' falls on the red wing of the C IV emission, is close to the systemic redshift and is partly overlapped with the Fraunhofer A band of O$_2$; the second, that we will call ``blue-NAL'', is overlapped with the BAL and is discussed in \citet{Sria00}. A continuum, extrapolated from two absorption-free regions bluewards of C {\scriptsize IV} emission (see section 4), is also shown. A gaussian fit to the emission line provides a width $\sigma_{C IV}=3450\pm 60$ km s$^{-1}$.

A further spectrum, of resolution $\lambda/\Delta \lambda \sim 700$, obtained at Telescopio Nazionale Galileo (TNG), La Palma, in 2011 (Piconcelli et al. 2013, in preparation) has also been included in the analysis.

\subsection{Photometry}
Photometry in the R band of APM 08279+5255 was obtained by \citet{Lewi99} on 23 epochs, between April 1998 and June 1999, with the main goal of studying the intrinsic or microlensing character of variability. They provided relative magnitudes, with respect to a reference star $S_2$, for the QSO and other 3 stars in the field, $S_1$, $S_3$, $S_4$ (see their Figure 1), for the purpose of choosing the best reference star. We also measured R band magnitude difference with respect to another star $S$, which we adopt as a reference for spectrophotometric measurements (see Section 3). Unfortunately this star is not included in the \citet{Lewi99} field. However their star $S_1$ is included in our field, so that we can rescale their photometry to ours, by defining:
\begin{eqnarray}
&\Delta R_{QS}=R_Q-R_2+\langle R_2-R_1 \rangle+\langle R_1-R_S \rangle&
\\
&\Delta R_{1S}=R_1-R_2+\langle R_2-R_1 \rangle+\langle R_1-R_S \rangle&
\nonumber
\end{eqnarray}
where $\langle R_2-R_1 \rangle$  and $\langle R_1-R_S \rangle$ are average values computed on the \citet{Lewi99} and our light curves respectively. Despite the larger uncertainty of $R_1$ as compared with those of $R_2$ or $R_S$, the uncertainty on the inter-calibration of the two data sets given by the average quantity $b=\langle R_2-R_1 \rangle+\langle R_1-R_S\rangle $ is $\sigma_b=[\sigma_{2-1}^2/(N_{2-1}-1)+\sigma_{1-S}^2/(N_{1-S}-1)]^{1/2} \sim 0.005$ mag, where $N_{1-2}=23$ and $N_{1-S}=24$ indicate the number of observations in the \citet{Lewi99}  and  our monitoring campaigns respectively. This quantity is smaller than the internal uncertainties of the individual data sets $\sigma_{1-2}\sim 0.023$ mag and $\sigma_{1-S}\sim 0.009$. The latter, in turn, can be assumed as conservative estimates of the uncertainty on the QSO photometry. In total we have at our disposal 49 photometric observations from 1998 to 2012  which can be correlated with spectral variations. The complete light curve is shown  in Figure \ref{6-w-r(t)}.
 
\section{Observations and data reduction}

In 2003 we started a spectrophotometric monitoring campaign  aimed at measuring the mass of four luminous QSOs  \citep{Trev07} by the reverberation mapping technique \citep{Blan82,Pete93}. Observations were carried out at the 1.82 m Copernicus telescope of the Asiago Observatory, equipped with the Faint Object Spectrograph \& Camera (AFOSC). Spectroscopic observations were carried out using a 8".44-wide slit and a grism with a dispersion of 4.99~\AA~ pixel$^{-1}$, providing a typical resolution of $\sim$ 15~\AA~ in the spectral range 3500-8500~\AA~ ($\lambda/\Delta \lambda \sim 400$). Exposures are carried out 
after orienting the slit to include both the QSO and a reference star $S$ of comparable magnitude ($R=14.4$), located at $\alpha$ 08 31 22.3  $\delta$ +52 44 58.6 (J2000), as internal spectrophotometric calibrator. The wide slit avoids both differential diffraction effects and possible different fractional losses of the QSO and the reference star, but limits the accuracy of the wavelength scale. In fact, the position of the QSO can fluctuate within the slit,  affecting both the line profile and the position on the wavelength scale. Thus, after the calibration with the spectral lamps, the $\lambda$ scale must be readjusted, and the entire procedure has a residual uncertainty  of  $\sim$ 3 \AA~ r.m.s. 
 
Typical spectroscopic observations consist of 2 subsequent exposure of 1800 s each, plus direct images in the R band to detect possible variations of the reference star with respect to other stars in the $\sim 8\times 8$ arcmin$^2$ field centred on the QSO. A journal of the observations is reported in Table \ref{tab-journal}, where the INT, Keck, Steward, HST and TNG observations are also indicated. In total we have 28 spectra from 1998 to 2012.

\begin{table}
\caption{List of the observations}            
\label{tab-journal}      
\centering                          
\begin{tabular}{l l l}        
\hline\hline                

  Date & MJD & Telescope \\    
\hline                        
  1998 Mar & 50875.7 & INT$^b$ \\
  1998 May & 50945.8$^a$ & Keck$^c$ \\
  1998 Nov & 51132.5 & Steward Obs.$^d$ \\
  2001 Dec  & 52270.4$^a$ & HST$^e$ \\
  2003 Feb & 52695.4 & Asiago Obs. \\
  2003 Nov & 52963.6 & '' \\
  2004 Feb & 53047.3 & '' \\
  2004 Feb & 53049.3 & '' \\
  2005 Jan & 53388.4 & '' \\
  2005 Apr & 53475.3 & '' \\
  2006 Mar & 53797.5 & '' \\
  2006 Nov & 54068.7 & '' \\
  2006 Dec & 54091.4 & '' \\
  2007 Feb & 54145.3 & '' \\
  2007 Apr & 54201.3 & '' \\
  2007 Dec & 54435.7 & '' \\
  2008 Jan & 54472.5 & '' \\
  2008 Feb & 54513.4 & '' \\
  2008 Dec & 54807.6 & '' \\
  2009 Jan & 54835.4 & '' \\
  2009 Feb & 54884.3 & '' \\
  2009 Mar & 54914.4 & '' \\
  2011 Apr & 55653.3 & '' \\
  2011 May & 55684.4 & TNG$^f$ \\
  2011 Nov & 55894.5 & Asiago Obs. \\
  2011 Dec & 55915.5 & '' \\
  2012 Feb & 55985.3 & '' \\
  2012 Nov & 56238.4 & '' \\
\hline                                  
\end{tabular}
\tablefoot{$^a$average of the observing MJD,
$^b$\citet{Irwi98},
$^c$\citet{Elli99},
$^d$\citet{Hine99},
$^e$\citet{Lewi02},
$^f$ Piconcelli et al. 2013 (in preparation)}.
\end{table}
The data reduction is described in \citet{Trev07} and briefly summarised below. The spectra $Q(\lambda)$ of the QSO and $S(\lambda)$ of the reference star were extracted with the standard IRAF procedures. The QSO/star ratio is computed for each exposure $k=1,2$:
\begin{equation}
\mu^{(k)}(\lambda)=Q^{(k)}(\lambda)/S^{(k)}(\lambda).
\end{equation}

This  quantity is independent from extinction changes during the night. This allows us
to check for consistency between the two exposures and to reject the data if inconsistencies occur (under the assumption that intrinsic QSO variations are negligible during 1 h observing time). In fact, typical spectra of two consecutive exposures have a ratio $\mu^{(1)}(\lambda)/\mu^{(2)}(\lambda)$ of order unity, with deviations smaller than 0.02 when averaged over 500 \AA, at least in the 4000 \AA-7000~\AA~ range. When discrepancies are larger than 0.04 both the exposures are rejected. A registration of the wavelength scale among different exposures is necessary due to small changes of the object position within the  wide slit (which are in general negligible in the case of pairs of consecutive exposures). Then, the QSO and star spectra taken in the two exposures are co-added and the ratio 
\begin{equation}
\mu_i(\lambda)=(Q_i^{(1)}+Q_i^{(2)})/(S_i^{(1)}+S_i^{(2)})
\end{equation}
is computed, at each epoch  $i$.

The reference star is flux calibrated  at a reference epoch, and the main telluric absorption features (Fraunhofer A and B bands $\lambda \lambda 7594, 6867$~\AA~ and water $ \lambda 7245$ \AA) are removed from the calibrated spectrum by interpolating the relevant absorption region with a spline function trough the spectral points in two intervals around each absorption feature. This provides us with a flux-calibrated absorption free spectrum of the reference star $f^S(\lambda)$. All the flux-calibrated quasar spectra are then obtained as:  $f^Q_i(\lambda) \equiv  \mu_i(\lambda) f^S(\lambda)$. Notice that the calibrated star spectrum $f^S(\lambda)$ adopted is the same for all epochs, thus it does not affect the {\it relative} flux changes. We stress also that the telluric absorption features are removed in this way since they do not affect the value of $\mu_i(\lambda)$. The same is obviously true for the differences in airmass and atmospheric extinction. 
\begin{figure}
   \centering
\resizebox{\hsize}{!}{\includegraphics{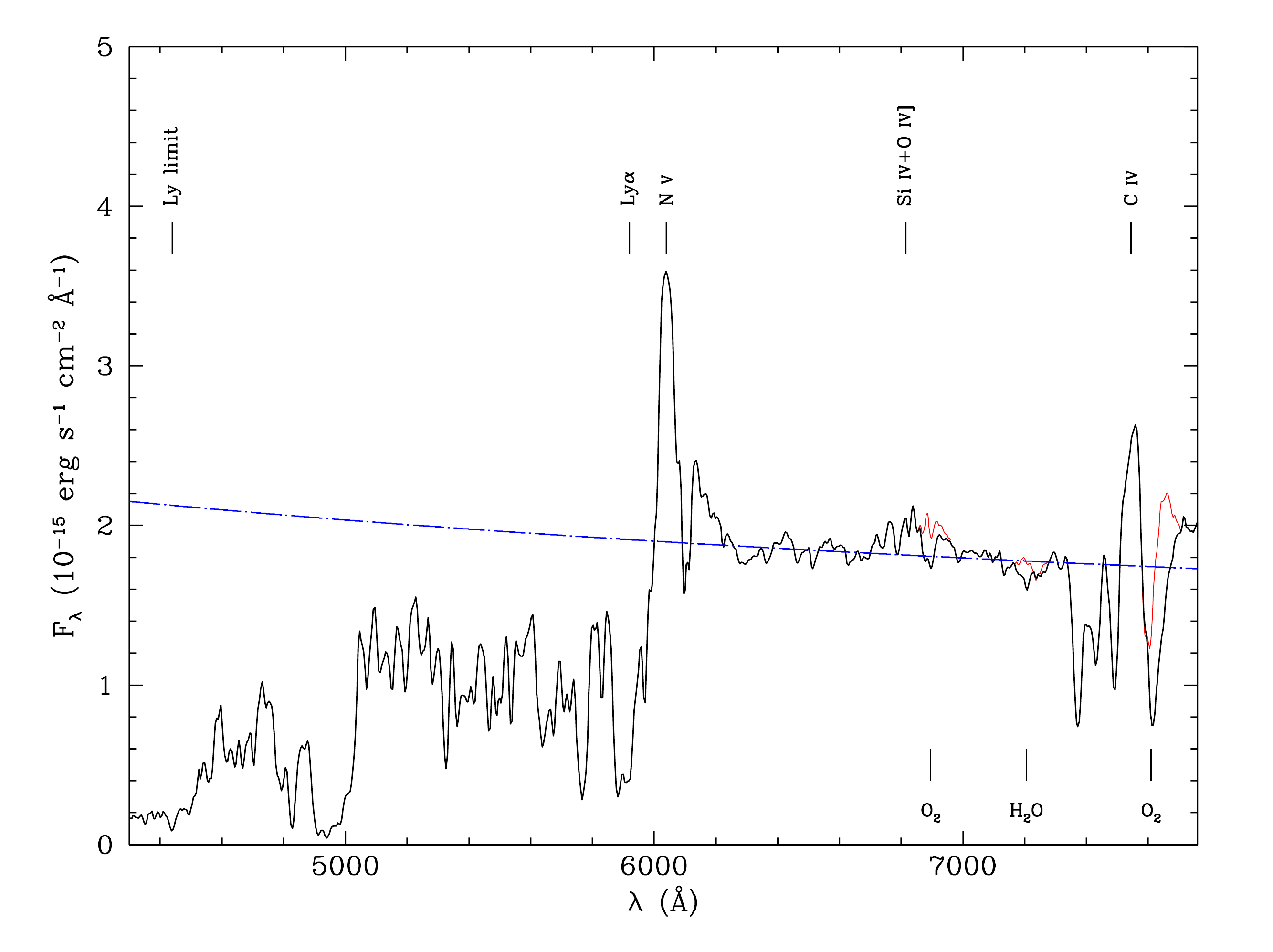}}
      \caption{
Spectrum of APM 08279+5255 obtained with the AFOSC camera at the 1.82 m Copernicus Telescope of the Asiago Observatory. The identification of the main absorption and emission features are indicated. The red line represents the correction for the telluric O$_2$ and H$_2$O bands. A power-law fit through spectral regions relatively free from absorption or emission features is also shown (see Sect. 4.1).
}
         \label{fig_2-sp-asia}
   \end{figure}
This is true, in particular, for the Fraunhofer A band which, in the case of APM 08279+5255,
falls just redwards of the C {\scriptsize IV} emission line and is also partially overlapped to the already mentioned C {\scriptsize IV} ``red-NAL'' at $z\sim 3.911$. An example of calibrated spectrum obtained at the Copernico telescope is shown in Figure \ref{fig_2-sp-asia}.
Thanks to the availability of the high resolution spectrum obtained  at Keck telescope by \citet{Elli99}, we checked that the procedure adopted to remove the telluric absorptions in low resolution spectra is satisfactory, at least to a first order.
This was checked by first   correcting for $O_2$ absorption  the high resolution spectrum, where the individual lines of the band are detected, using the spectrum of the calibration star  (Feige 34). Then we smoothed the corrected spectrum,   down  to the  resolution  $\lambda/\Delta \lambda \sim 400$ of our spectra. 
For comparison, we  smoothed  down to the same resolution both the QSO and calibration star spectra, without correcting for O$_2$ absorption. We then applied the procedure adopted on our low resolution data to remove telluric absorptions. The equivalent width of the C IV absorption feature, computed in both cases, appears consistent within 5\%, despite the Keck spectra of the QSO and the calibration star are not taken simultaneously, as instead is the case for our  QSO and the reference star $S$. 

\section{Spectral variability}
\subsection{The BAL and NAL equivalent width variability}

 \begin{figure}
    \centering
   \resizebox{\hsize}{!}{\includegraphics{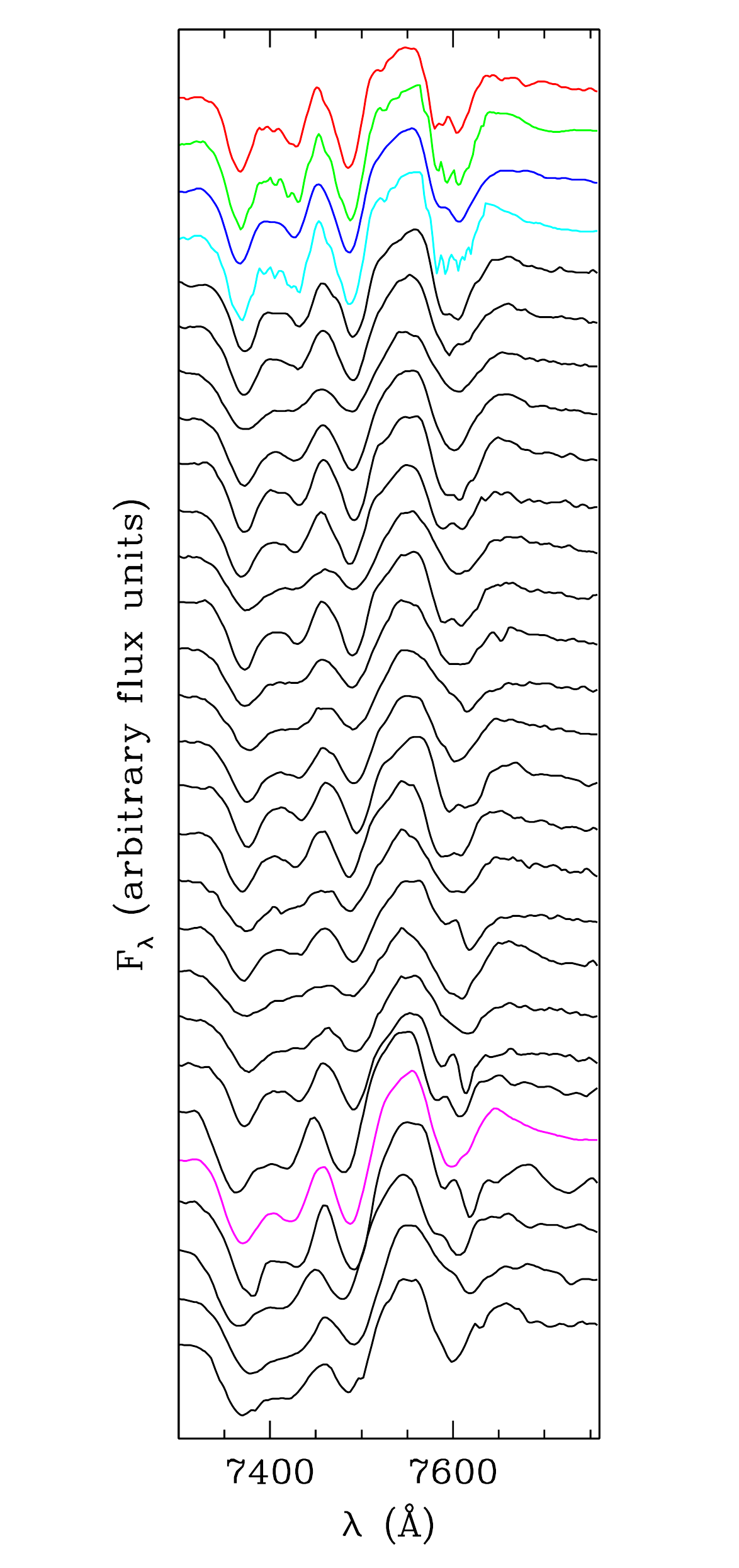}}
      \caption{The evolution in time of the spectral region around C {\scriptsize IV} emission line.  From top to bottom, the first (red) and third (blue) spectra correspond to INT and Steward Observatory data, the second (green), the fourth (cyan) and 24-th (magenta) epochs correspond respectively to Keck, HST and TNG spectra, smoothed to match the resolution of the Asiago spectra (see Table 1).  All spectra are shown with the same flux scale and an arbitrary shift between each other. At a glance it is clear that the BAL feature retains its global {\it shape} during time.}
   \label{fig_3-sp-bal(t)}
   \end{figure}  
   
The present analysis is focused on the spectral region around C {\scriptsize IV} emission, whose evolution in time is shown in Figure \ref{fig_3-sp-bal(t)}. 

We stress that at the time of discovery the systemic redshift was assumed $z$=3.87 and was later revised to $z$=3.911, as measured by \citet{Down99} from CO(4-3) and CO(9-8) molecular lines at $\lambda$650 $\mu$m and $\lambda$290 $\mu$m, implying a Doppler blueshift velocity of 2500 km s$^{-1}$ for the emission lines. While \citet{Lund07} include in their sample only BALs separated from the associated emission peak by more than 3600 km s$^{-1}$, on the contrary we are forced to consider the absorption occurring on the blue wing of the emission line, since in our case the BAL is separated from the peak by less than 800 km s$^{-1}$. At each epoch, we compute a continuum  as a power law fit through the observed spectrum, in two wavelength intervals which are free from major emission or absorption features. The rest-frame regions $\lambda\lambda 1250-1370$~\AA~ and $\lambda\lambda 1440-1470$~\AA~ were selected following \citet{Cape11} and \citet{Barl92} respectively, for consistency \citep[but see also][]{Vand01}.
The C {\scriptsize IV} emission line profile is strongly affected by the BAL, on the blue side, and by the Fraunhofer A band plus the red-NAL system on the red side. However, we can use both the Keck and HST spectra to identify the unabsorbed wavelength intervals around C {\scriptsize IV} and fit the emission line profile.

We fit the emission line  with a gaussian profile, keeping the central wavelength fixed to the value corresponding to $z=3.87$, deduced from the  Si {\scriptsize IV}+O {\scriptsize IV]}, C {\scriptsize III]} and  N {\scriptsize V} \citep{Irwi98}, and the line width to the value $\sigma_{C IV}=3450\pm 60$ km s$^{-1}$ (see Sect. 2), letting the line amplitude be the sole varying parameter.
In doing this, we are assuming that relative amplitude changes are larger than the relative changes in the shape of the line. We can now consider a pseudo-continuum consisting in a proper combination of continuum and emission line fluxes. The spatial location and the size of the absorbing gas is unknown and it could cover only part of the continuum and/or emission line sources. Thus we can write the observed flux as:
\begin{equation} 
f(\lambda)=f_c (1-\xi_c)+f_l (1-\xi_l)+(\xi_c  f_c+\xi_l  f_l) e^{-\tau(\lambda)}
\end{equation}
where $f_c, \xi_c, f_l, \xi_l$ indicate the continuum and emission line fluxes and covering factors, and $\tau(\lambda)$ is the optical depth which, in a simple model, is assumed to be the same for photons emitted from the continuum source and the emitting clouds. Since the size of the continuum-emitting region is expected to be much smaller than the Broad Line Region (BLR) size, then $\xi_l \le \xi_c$. In the following we will assume $\xi_c=1$. Moreover, the high resolution Keck spectrum shows that some  individual absorption features in the BAL have a residual flux which is smaller than the  emission line flux at the  same wavelenght (see Figure \ref{fig_1-sp-keck-hst}), implying that the pseudo-continuum, which is absorbed by the outflowing gas, must contain a contribution from the emission line corresponding to $\xi_l \ge (1- f/f_l) \approx 0.3$. 
Thus  we compute the equivalent widths  $W$:
\begin{equation}
W(t)=\int{}{}\frac{f_c+f_l-f}{\xi_c f_c+\xi_l f_l}d\lambda
\end{equation}
according  to the two extreme  assumptions $\xi_l=1$ and $\xi_l=0.3$, the former corresponding to the pseudo-continuum adopted by \citet{Cape11}. The values of $W(t)$ computed with $\xi_l=0.3$ are a factor $\approx 1.2$ larger than for $\xi_l=1$. This factor, however, is virtually constant in time and thus does not affect the following discussion of the time behaviour of the absorption. Thus we adopt $\xi_l=1$ henceforth.

\begin{figure}
    \centering
   \resizebox{\hsize}{!}{\includegraphics{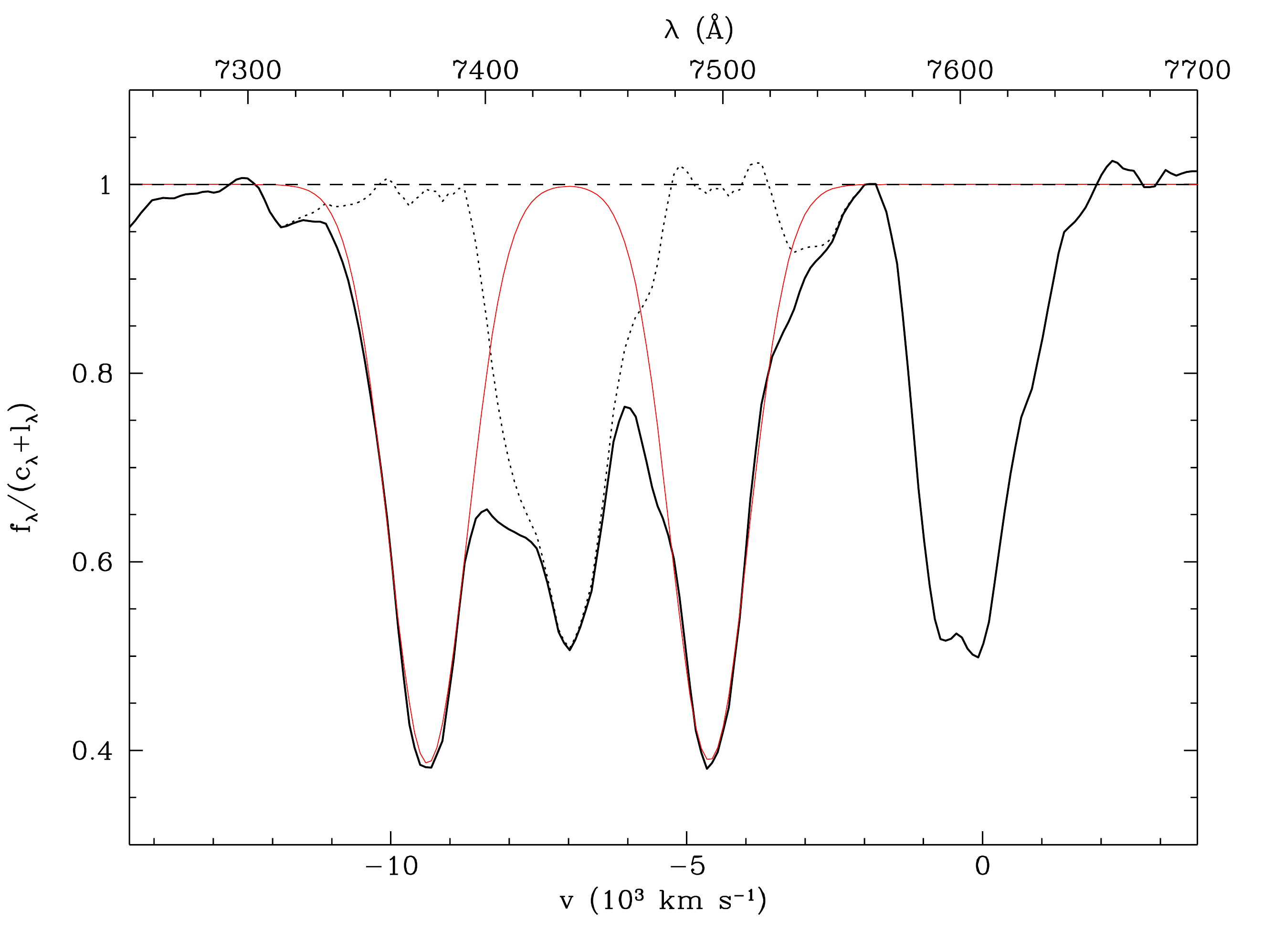}}
      \caption{ The decomposition of the C {\scriptsize IV} absorption structure at a particular epoch (MJD 52695.4). The flux is divided by the pseudo-continuum computed with  $\xi_c=\xi_l$=1. Two Gaussian profiles (red) represent the main component  $BAL_1$ and $BAL_2$ of the BAL.  The dotted line is the residual absorption corresponding to the blue NAL. The velocity scale is based on the systemic redshift $z$=3.911.
      }
   \label{4-gauss-fit}
   \end{figure}

The uncertainty on the equivalent width $W_i$ at the $i$-th epoch is estimated as follows. Since in our monitoring we have two exposures of 1800 s for each spectrum, we compute the difference $\delta W_i=|W^{(2)}_i-W^{(1)}_i|$ between the equivalent width as computed with each exposure. They appear correlated with the equivalent widths $W_i$. Thus we assume constant fractional uncertainty $\varepsilon=(\langle(\delta W/W)^2\rangle)^{\frac{1}{2}}$, where  the angular brackets indicate the average on the entire set of measurements, and we adopt as r.m.s. error $\sigma_W= \varepsilon W$, where $\varepsilon =0.06$. From Figure \ref{fig_3-sp-bal(t)} it appears that the general shape of the absorbing systems is conserved in time, as stated quantitatively in the following. High resolution spectra show that the absorption bluewards of C {\scriptsize IV} emission consists of two main broad structures around -10000 and -4500 km s$^{-1}$ respectively (see Figure \ref{fig_1-sp-keck-hst}), and the system of narrow absorption lines, between -9000 and -7000 km s$^{-1}$ \citep{Sria00} which we call ``blue-NAL''. In our low resolution spectra (see Figure \ref{4-gauss-fit}) we decompose the absorption profile by fitting 2 Gaussians, in the spectral regions $\lambda\lambda 7300-7390$~\AA~ and $\lambda\lambda 7470-7510$~\AA~ respectively,   corresponding to the  two broad structures (see Figure \ref{fig_1-sp-keck-hst}).  A third structure, corresponding to the ``blue-NAL'' is obtained as the residual spectrum after the two gaussian components are subtracted. The Gaussians have no physical meaning, but they are simply used to describe apparent structures at the resolution of our spectra. Of course the ``blue-NAL'' residual might be affected by an unknown contribution from the broad absorption.

The central wavelengths $\lambda_1$ and  $\lambda_2$, the widths $\sigma_1$ and $\sigma_2$ and the amplitudes of both the gaussian components are free parameters in the fit performed at each epoch.  
The average central wavelengths of the two BAL features are $\bar{\lambda}_1=7373$ \AA,  $\bar{\lambda}_2=7489$~\AA~ and the relevant standard deviation are $\Sigma_{\lambda_1}=\Sigma_{\lambda_2}=3$ \AA.
\begin{figure}
    \centering
   \resizebox{\hsize}{!}{\includegraphics{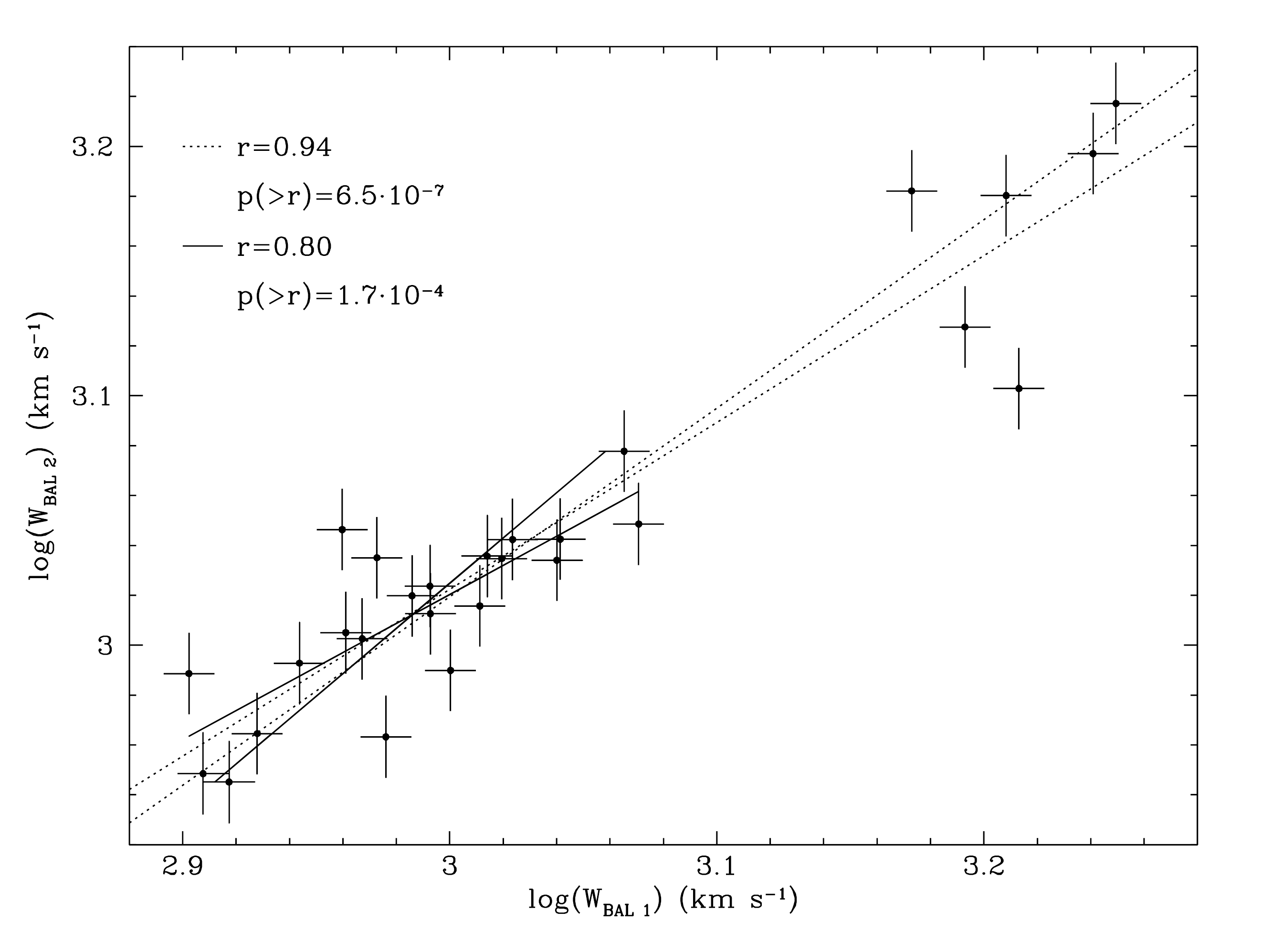}}
      \caption{ Logarithms of the equivalent widths $W_{BAL_1}$   and $W_{BAL_2}$  expressed in km s$^{-1}$. The continuous and dotted pairs of regression lines (one component versus the other and vice versa) are computed respectively excluding or including the observations at the last 6 epochs. The correlation coefficients and the relevant probabilities of the null hypothesis are indicated. The correlation is high and statistically significant also when the last 6 points are excluded.}
      
   \label{corre-compo}
   \end{figure}
The central wavelengths of the BAL components show small random shifts, comparable with  the uncertainty of our wavelength scale. Instead, the amplitudes of both the components undergo significant changes.
However the variations of their equivalent widths $W_{BAL_1}$ and $W_{BAL_2}$ are strongly correlated, $r_{1,2}=0.94$, $P(>r_{1,2})=6.5\cdot 10^{-7}$, as shown in  Figure \ref{corre-compo} where the equivalent widths are computed in units of velocity with respect to the emission line, for consistency with the literature.
This suggests to consider $W_{BAL}=W_{BAL_1}+W_{BAL_2}$ as representative of the total  BAL absorption,
 in the subsequent variability analysis.  Notice that this behaviour is completely different from that of FBQS J1408+3054 which exhibits strong changes of the absorption shape, suggesting the motion of part of the absorbing clouds out of the line of sight \citep{Hall11}.
 \begin{figure}
    \centering
   \resizebox{\hsize}{!}{\includegraphics{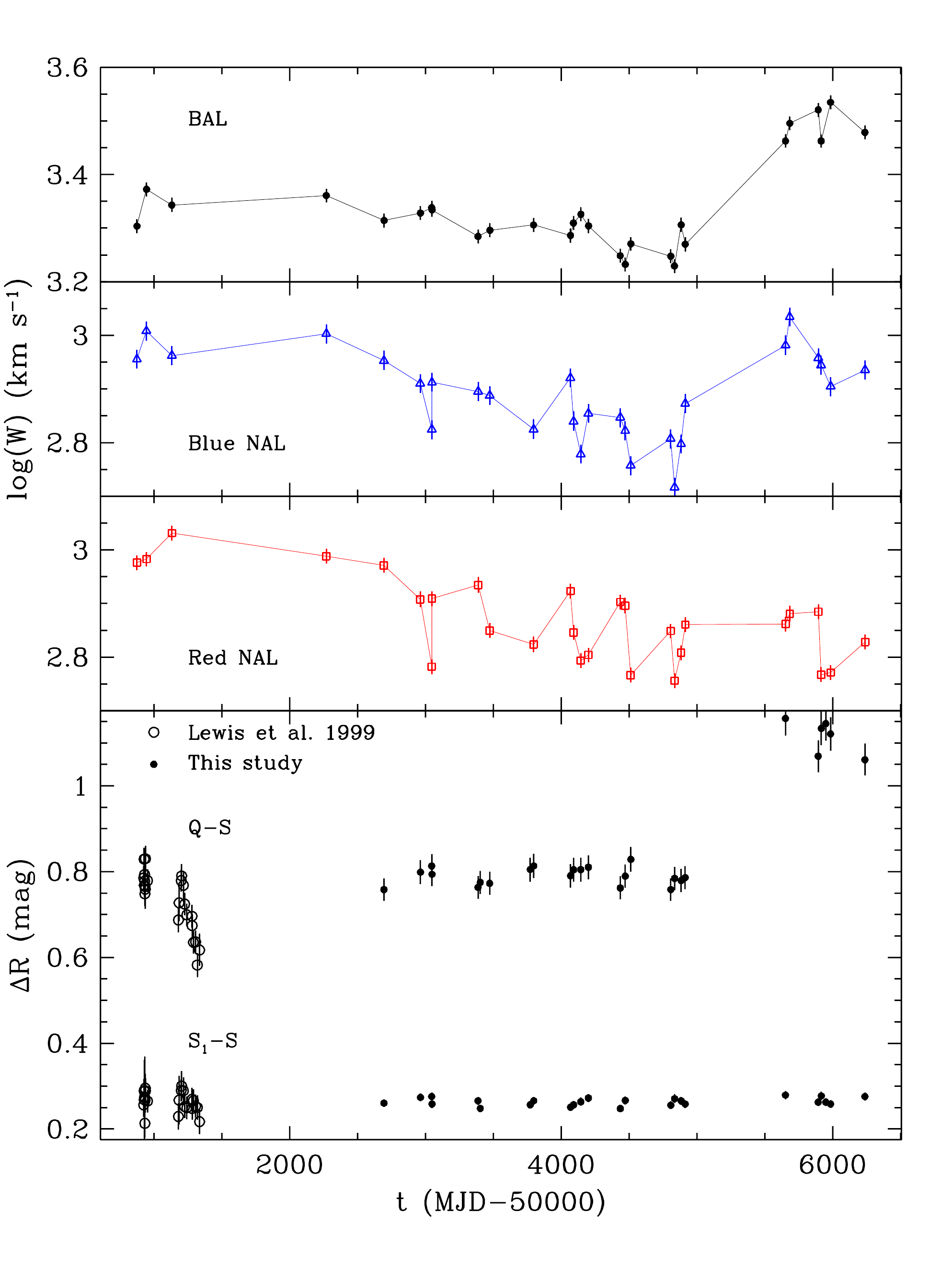}}
      \caption{ {\it Upper panels}: the equivalent widths $W_{BAL}$, $W_{NALb}$ and $W_{NALr}$ of the C {\scriptsize IV} absorptions as a function of time.   {\it Bottom panel}: the R band light  curve $\Delta R_{QS}$ of APM 08279+5255 normalised to the reference star $S$  adopted in the Asiago monitoring campaign (upper curve). For comparison the magnitude difference $\Delta R_{1S}$ between the star $S_1$ of  and our reference star $S$ is also reported  as an indication of the photometric accuracy. Photometry, relative to the star $S$, is computed from magnitude differences for our data, while it is recalculated according to eq. (1) for \citet{Lewi99} data.
      }
   \label{6-w-r(t)}
   \end{figure}   
For  the red and blue NAL systems respectively,  we compute the relevant equivalent widths $W_{NALr}$  and $W_{NALb}$ by simply integrating the residuals in the ranges $7550-7660$~\AA~ and $7380-7480$~\AA.
Figure \ref{6-w-r(t)} shows $W_{BAL}$,  $W_{NALr}$  and $W_{NALb}$  as a function of time $t=$MJD$-50000$.  All of these quantities show  a slow decreasing trend from  $t \sim 1000 $ to  $t \sim 5000$. Subsequently the $W_{BAL}$ undergoes a sudden increase, while  $W_{NALr}$ continues the decreasing trend. 
$W_{NALb}$ shows an intermediate behaviour, but it might be affected by the BAL absorption. 
High resolution monitoring is necessary to disentangle BAL and NAL absorption variations in this spectral region. 

The overall appearance of the BAL structure is relatively stable (two minima, with no significant wavelength variation, and amplitudes varying proportionally), while its equivalent width changes significantly. This  suggests that the global properties of the density and velocity fields of the absorbing gas remain unchanged during variations of $W(t)$. 
The simplest explanation of such a behaviour is a change in the ionisation status of the absorbing gas.
This conclusion is independent of the correlation of $W(t)$ with the variations of the $R$ band continuum, which is discussed in the next subsection. In fact, the R band measures the continuum at $\lambda \approx 1300$~ \AA, while the ionising continuum, relative to the C$^{3+}\rightarrow$ C$^{4+}$ transition, corresponds to $\lambda_{3,4} \approx 200$~\AA, and the two continua are not necessarily correlated.
 However the shape of the BAL structure is not strictly constant, since the average slope in Figure  \ref{corre-compo}  is about 0.7, implying a decrease of the ratio $W_{BAL_2}/W_{BAL_1}$  for increasing $W_{BAL}$. This will be discussed in section 5.

\subsection{Correlation analysis}


\begin{figure*}
   \centering
   	\resizebox{\hsize}{!}{\includegraphics{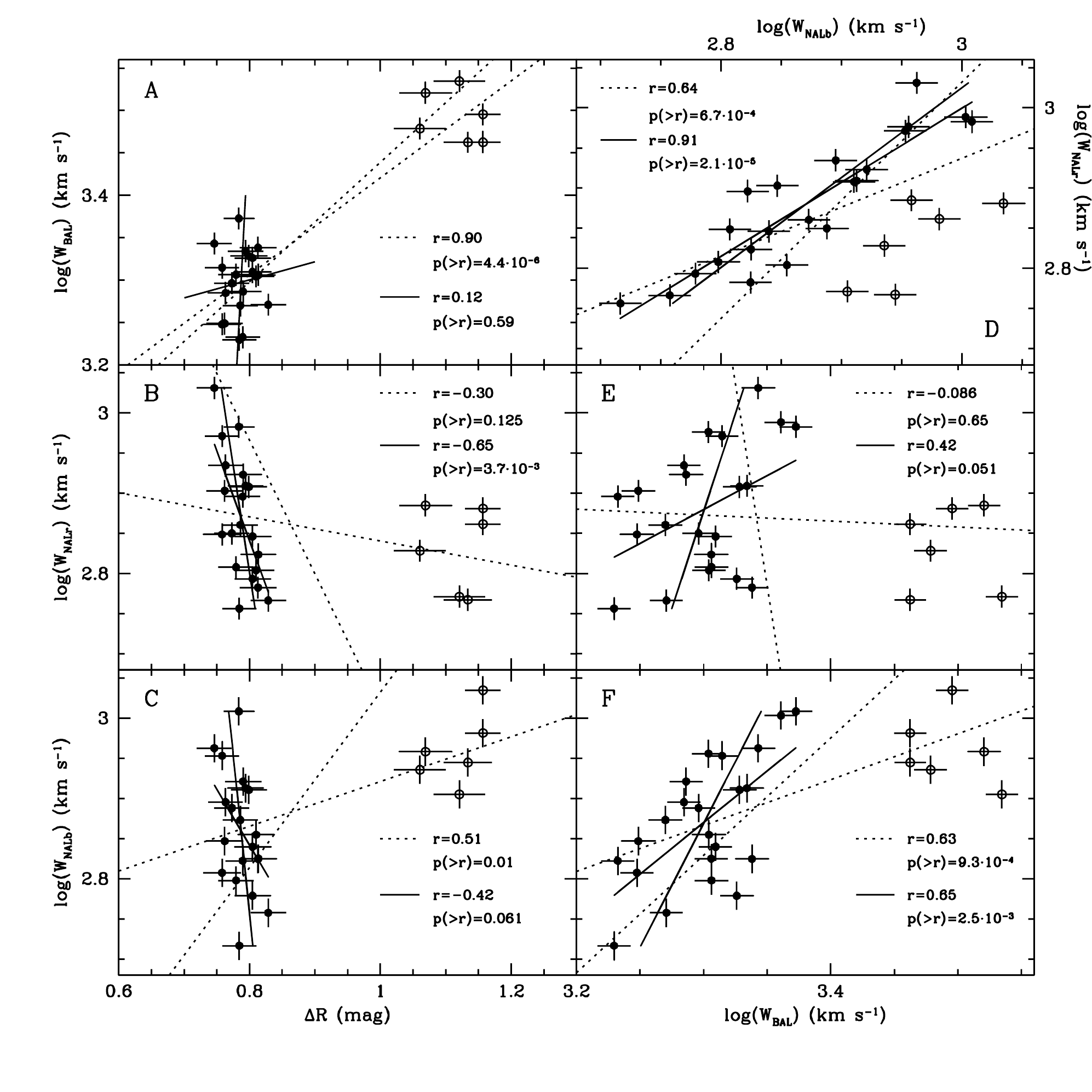}}
   \caption{ The six scatter plots for the four quantities  $W_{BAL}$, $W_{NALr}$, $W_{NALb}$ and $\Delta R$:
   {\it (A)}$W_{BAL}$ vs.  $\Delta R$; 
   {\it (B)}$W_{NALr}$ vs. $\Delta R$;
   {\it (C)}$W_{NALb}$ vs. $\Delta R$;
   {\it D)}$W_{NALr}$ vs.  $W_{NALb}$;
   {\it (E)}$W_{NALr}$ vs. $W_{BAL}$;
   {\it (F)}$W_{NALb}$ vs. $W_{BAL}$.
   Filled circles represent data for  MJD $<$ 55000, open circles for MJD $>$ 55000 (the last six epochs). Pairs of regression lines  represent linear fits of one variable versus the other and vice versa. Dotted lines refer to the whole set of data, while continuous lines refer to filled circles only. Correlation coefficients and the relevant probabilities are indicated in each panel.
   }
  \label{corre-r-w-nal}
\end{figure*}

So far, we have not considered  the relation between  the  absorptions  changes and  the  source brightness variations $\Delta R$ of the $R$ magnitude. For the four quantities  $W_{BAL}$, $W_{NALr}$, $W_{NALb}$ and $\Delta R$, we can analyse 6 scatter plots, shown in Figure \ref{corre-r-w-nal}, the relevant correlation coefficients and probabilities of the null hypothesis.  We comment them in turn.\\
{\it a)} $W_{BAL}$ vs. $\Delta R$. These quantities show a strong and significant correlation, $r=0.9$ and $P(>r)=4.4\cdot 10^{-6}$. This is the first time, to our knowledge, that a continuum-equivalent width correlation is demonstrated directly in an individual BAL quasar. Some previous evidences were derived as a possible  interpretation of the ensemble analysis of BAL variability. In fact, \citet{Barl93} found that most variable BALs tend to occur in objects with at least some  broadband changes, suggesting some correlation. It is important to notice that, if we exclude the last 6 epochs, the correlation drops to $r=0.12$, this  is due to the fact that during the slow decrease of $W_{BAL}$, the $R$ continuum stays almost constant and the correlation is entirely due to the fast increase of $W_{BAL}$ when the $R$ flux decreases significantly.\\
{\it b)} $W_{NALr}$ vs. $\Delta R$. At variance with $W_{BAL}$, the red-NAL equivalent width $W_{NALr}$ is only marginally anti-correlated with $R$, if we consider all the observations. However, this negative correlation becomes more significant if we exclude the last 6 epochs where there is a significant variation of $R$.\\
{\it c)} $W_{NALb}$ vs. $\Delta R$.
 The exclusion of the last 6 epochs corresponds to a negative correlation but only marginal. At variance  with {\it b)}, the points corresponding to the last 6 epochs lie in the top right of the panel, implying an even smaller correlation when all epochs are considered. This, however, is likely due to the ``contamination'' of the residual corresponding to the blue NAL, by the absorption of the underlying BAL.\\
{\it  d)} $W_{NALr}$ vs. $W_{NALb}$. A moderate correlation is found, entirely due to the lack of points in the lower right corner. However, if we exclude the last 6 epochs, the correlation becomes higher and more significant. Again the fact that the  last five epochs do not follow the general trend can be explained by the contamination of $W_{NALb}$ due to BAL absorption.\\
{\it e)} $W_{NALr}$ vs. $W_{BAL}$. There is no correlation if we include all the epochs.
Excluding the last 6 epochs, $W_{NALr}$ and $W_{BAL}$ show a marginal positive correlation.  We stress that this happens despite $W_{BAL}$ is not correlated with $R$, namely the correlation  of $W_{NALr}$ and $W_{BAL}$ is independent of their relation with $R$. This suggests the possibility  that both $W_{NALr}$ and $W_{BAL}$ are influenced by the variation of the same ionising continuum \citep[see][]{Hama11}.\\
{\it f)} $W_{NALb}$ vs. $W_{BAL}$. The situation is similar to the previous one, except that the points corresponding to the last  6 epochs lie on the top right corner. This makes the correlation positive also if all the epochs are considered. We notice, however, that this can be explained by the ``contamination'' of  $W_{NALb}$ by the BAL absorption, as in {\it c)}.\\ 
 
We notice that  the blue NAL, which covers roughly the interval from -8500  km s$^{-1}$ to -6000 km s$^{-1}$ lies in the middle of the ejection velocity range of the BAL. This may mean  that this NAL is caused by clouds embedded in the BAL outflow, though of course the sole velocity is not sufficient to know whether or not the absorbing clouds are spatially located inside the outflowing BAL gas. Its column density is smaller than the column density of the red NAL, as can be seen from the high resolution Keck spectrum (see Figure \ref{fig_1-sp-keck-hst}). The maxima between the absorption line of the blue NAL do not reach neither the flux of the emission line nor the continuum, suggesting the presence of a contribution of  an underlying BAL absorption. The minimal flux is about one tenth of the continuum. On the other hand the velocities of the red NAL features lie roughly within a 1000 km s$^{-1}$  around the systemic velocity. The minima are saturated and the to maxima are close to the  pseudo-continuum, implying a covering factor $\xi_l  \approx 1$ for this NAL. Thus the physical conditions of red and blue NALs are different. However, it is difficult to establish whether or not  the differences found in the correlations can  be associated to physical differences because of the contamination of the blue NAL caused by the BAL, which would require high resolution variability studies.

 \subsection{Structure function analysis}
\begin{figure}
    \centering
   \resizebox{\hsize}{!}{\includegraphics{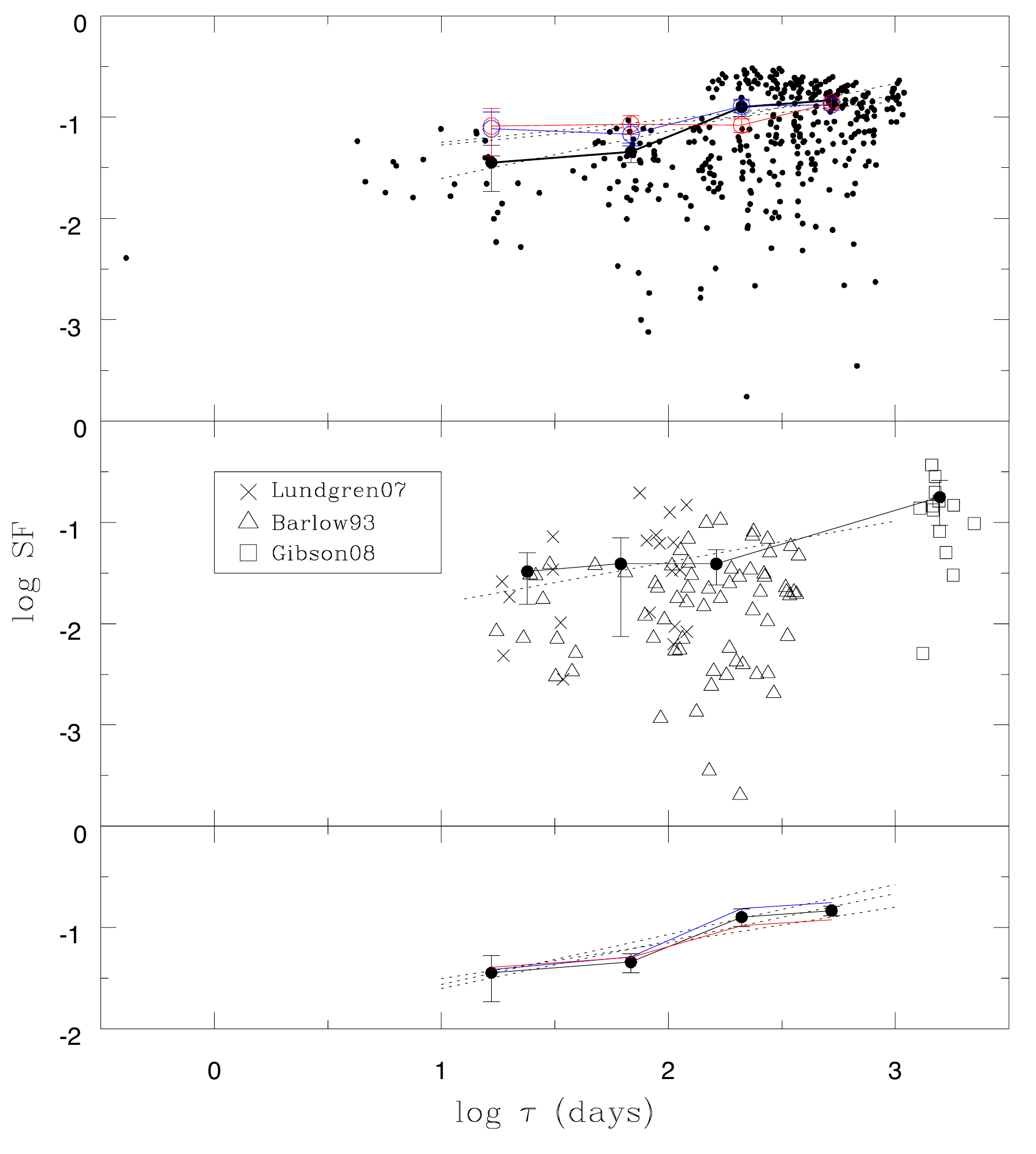}}
      \caption{ {\it Top panel }: structure function of APM~08279+5255; small dots represent the UDSF points (eq. 6) relative to BAL variations; filled circles connected with continuous line represent the corresponding binned SF; blue and red open circles connected by continuous lines represent the binned SF for blue and red NALs (blue and red respectively). The relevant fitting relations corresponding to Table \ref{tab-fit} are shown as dotted lines. {\it Middle panel }: ensemble structure function based on data from the literature \citep{Barl93,Lund07,Gibs08}; UDSFs  with different symbols as specified in the inset; filled circles connected by continuous line represent the binned ensemble SF and dotted line shows the best fit relation of Table \ref{tab-fit}.  {\it Bottom panel}: structure functions of the two BAL components  of APM~08279+5255;  BAL$_1$ blue line, BAL$_2$ red line, for comparison  black filled circles and the connecting line show the structure function of the total W$_{BAL}$ as in the top panel. }
   \label{8-sf}
   \end{figure}

Our unprecedented temporal sampling of the spectral variability of a BAL quasar, allows us to apply a structure function (SF) analysis  commonly adopted in the study of AGN flux variability in the radio \citep{Hugh92, Hufn92},  optical  \citep{Trev94,Vand04,Wilh08,MacL12}  and X-ray bands \citep{Fior98,Vagn11}.
In practice, with $N=28$ observations we can compute $N(N-1)/2=378$ values of the unbinned discrete structure function \citep{di-C96,Trev07}:
\begin{equation}
UDSF(\tau_{ij})= \sqrt{\frac {\pi}{2}}\left | log[W(t_i)]-log[W(t_j)] \right |, 
\end{equation}
\noindent
where $W$ may indicate $W_{BAL}$, $W_{NALr}$ or $W_{NALb}$ in turn, $t_i$ and $t_j$ are two observation epochs and $\tau_{ij}=(t_i-t_j)/(1+z)$ is the rest-frame time delay of the QSO at redshit z. The (binned) structure function can be defined in $N_{bin}$ bins of time delay,  centered at $\tau_k$, $k=1,N_{bin}$
\begin{equation}
SF(\tau_k)=\frac{1}{M_k}\left[\sum_{i,j} UDSF(\tau_{ij})\right], 
\end{equation}
\noindent
where the sum is extended to all the $M_k$ values of UDSF belonging to the $k$-th bin of  time delay. We refer to the logarithm of $W$ for analogy with the studies of flux variability which are commonly analysed in magnitude or logarithm of the flux. It is possible to define an {\it ensemble} SF by co-adding in each bin the UDSFs derived from a sample of objects. One advantage of this procedure is that  it can be applied even with only one pair of observations per object \citep[see e.g.][]{Vand04}, provided that the number of objects in the sample is large enough. It should be noted that, thanks to the  redshift distribution of the sources,   the entire range of rest-frame delays may be well sampled. In the  top panel of Figure \ref{8-sf} we show the UDSF for the BAL  of APM 08279+5255. For comparison, in the middle panel we show the {\it ensemble} SF of the BAL equivalent width variability obtained using the data available from the literature \citep{Barl93, Lund07, Gibs08}. This corresponds to Figure 12 of \citet{Gibs08}, in a slightly different presentation. 

At a first glance it appears that the unbinned distributions of UDSF points in the two panels  are similar. The binned SF for the BALs in the two panels have also similar amplitudes and slopes. Notice that, thanks to the high redshift of APM~08279+5255, time lags below 10 days, down to 0.4 days in the rest-frame, are sampled. In the top panel the binned SFs of the red and blue NAL are also reported. They look slightly flatter than the BAL SF due to larger variability at short time lags, which however could be simply due to a larger noise. In the bottom panel we show the SFs of the two broad absorption components BAL$_1$ and BAL$_2$. It appears that BAL$_1$ varies slightly more than BAL$_2$. This corresponds to about $\pm 20$\% difference with respect to the total BAL, in the bin of longest delay. This is clearly a consequence of the high correlation between the amplitude of the two components, with $W_{BAL_2}/W_{BAL_1}$ decreasing for increasing $W_{BAL}$, and vice versa (see Figure \ref{corre-compo}).
Amplitudes and slopes of a power-law fit, $\log SF= A \log \tau + B$, are reported in Table \ref{tab-fit}.

\begin{table}[h]
\caption{Best fit parameters of the Structure Functions }             
\label{tab-fit}      
\centering                          
\begin{tabular}{l c c  }        
\hline\hline                 

&  $A$  & $B$ \\    
\hline                        
APM 08279  $BAL$  &   0.47 $\pm$ 0.14 &  -2.07 $\pm$ 0.34  \\   
APM 08279 $NALb$ & 0.23 $\pm$ 0.10 & -1.47 $\pm$ 0.25 \\
APM 08279 $NALr$ & 0.22 $\pm$ 0.13 & -1.49 $\pm$ 0.30 \\
Ensemble $BAL$  & 0.41 $\pm$ 0.16 &  -2.20 $\pm$ 0.36 \\
APM 08279 $BAL_1$ & 0.49 $\pm$ 0.15 & -2.06 $\pm$ 0.37 \\
APM 08279 $BAL_2$ & 0.35 $\pm$ 0.09 & -1.86 $\pm$ 0.22 \\
\hline                                   
\end{tabular}
\tablefoot{According to the relation  $\log SF= A \log \tau + B$, with $\tau$ in days.}
\end{table}
It is interesting to compare these slope values with typical values found for optical and UV variability, where the slopes are in the range 0.3 to 0.45 \citep{Vand04,Wilh08,Baue09,MacL12,Wels11}. If the absorption variations that we are measuring are in fact due to a change in C$^{3+}$ column density, caused by variations of the ionisation status, we might be observing indirectly the structure function of the $200$~\AA~ variability. We notice, for completeness, that we have included in the ensemble structure function analysis only equivalent widths larger than 5~\AA~ on average, as done by \citet{Gibs08} so that the middle panel of our Figure \ref{8-sf} is essentially equal to their figure 12, except for the change of the variable reported in the ordinates. If instead we include all the points with smaller equivalent width, the SF slope becomes flatter, $A=0.25$, providing a hint of how the segregation or incompleteness of the data may affect the results. Nonetheless, we can say that, to a first order, the ensemble SF is similar to the SF of APM~08279+5255.

\section{Discussion}

We can summarise our results as follows.\\
      a){~The series of 28 spectroscopic observations spanning a 15 years time base shows that the BAL feature preserves its overall shape, with the two main components varying in a higly correlated way.}\\
       b){~In spite of this stability, both the BAL and the NAL features exhibit a significant variation in equivalent width.}\\
        c){~These two results suggest that both density and velocity fields of the absorbing gas remain stable.}\\
        d){~The R band magnitude shows no systematic trends until MJD $\sim$ 55000 but only small oscillations around the mean value, then it increases significantly ($\sim $ 30\% flux decrease) in the last 6 epochs.\\
        e){~The  flux decrease in the $R$ band is accompanied by an increase of the BAL absorption; a correlation analysis shows that for MJD $<$ 55000 there is no correlation between $W_{BAL}$ and $R$, but the simultaneous and significant increase of  $W_{BAL}$ and $R$ after that date determines a statistically significant correlation, $r=0.9$, $P(>r)=4.4\cdot 10^{-6}$ (panel A in Figure 7). Some indications of a possible correlation were suggested already by \citet{Barl93}, who noticed that ``most variable BALs tend to occur  in objects that show at least some broadband changes'', however we stress that it is the first time, to our knowledge, that a direct evidence of significant correlation is found}.\\
                     f){~$W_{NALb}$ and $W_{NALr}$ appear correlated, and in particular the correlation and its significance increase considerably ($r=0.9$, $P(>r)=2 \cdot 10^{-5}$) when the last 6 epochs are excluded. A straightforward interpretation is that the scatter of the points in the last epochs is due to a contamination of the blue NAL from the BAL, while intrinsically the red and blue NALs are strongly correlated. We do not comment further the correlations of $W_{NALb}$ with the other quantities because disentangling BAL and blue NAL variations would require new high resolution data.}\\   
     g){~The SF of the BAL variations in APM~08279+5255  has similar slope and amplitude to the ensemble
     structure function of BAL variations derived from a joint analysis of the data by \citet{Barl93}, \citet{Lund07}, and \citet{Gibs08}, and shown in the middle panel of Figure \ref{8-sf}.\\
      h){~The SFs of  red and blue NALs   of APM~08279+5255 have  similar slopes, smaller than BAL.}\\
       i){~ A comparison of the SFs of the two components of APM~08279+5255 BAL indicates a stronger variability of the one with higher ejection velocity.}\\
       j){ ~The SF slope  of APM~08279+5255 BAL is similar to typical values found for the optical/UV variability of  normal quasars \citep[e.g.][]{MacL12}.}

The above results might be easily explained  by changes in the ionisation status of the absorbing gas. 
The different amplitude of  $W_{BAL_1}$ and $W_{BAL_2}$ variations, which are strongly correlated, might mean that the two components have different levels of saturation. Under the assumption that absorption variability is driven by changes of the ionising flux ($\lambda \sim 200$ \AA),  the comparison of $W_{BAL}$ and $R$ changes, before and after MJD $\sim$ 55000, would correspond to a change of the average continuum slope between 200 \AA\ and 1500 \AA, indicating different physical conditions in the two epochs. This could happen, for instance, if before MJD $\sim$ 55000
the ionisation changes are due to small variations of the intensity of the ionising source,  which can be identified with the inner part of the accretion disk. The continuum at 1500 \AA\ could be substantially unaffected.  After MJD $\sim$ 55000, when a major variation occurs,  an absorbing cloud, crossing the line of sight between the ionising continuum source and the gas outflow responsible for the BAL, could reduce at the same time both the ionising continuum and the continuum at 1500 \AA, thus causing the observed correlation between $W_{BAL}$ and $R$.

The sudden increase of $W_{BAL}$ and $R$ after MJD $\sim$ 55000 does not occur for $W_{NALr}$.  Could this different behaviour be due to a delay of NAL variations? At variance with the emission lines, for absorption lines there is no delay caused by the geometry since the absorber lies along the line of sight of continuum photons \citep{Barl93}. Nonetheless, the intrinsic delay due to the recombination time of C$^{4+}$ atoms may be significant.  Assuming that the different behaviour of $W_{NALr}$ and $W_{BAL}$ is simply due to the fact that we do not see yet the  $W_{NALr}$  increase, we can infer an upper limit on the electron density $n_e \la 1/ (\Delta t \alpha_r) \approx 2 \cdot 10^4$ cm$^{-3}$, adopting a minimum (rest-frame) delay $\Delta t\sim$ 200 days and $\alpha_r=2.8 \cdot 10^{-12}$ cm$^3$ s$^{-1}$ \citep{Arna85}. This density is comparable to typical Narrow Line Region densities, thus suggesting this as a possible location of the absorbers. This is also consistent with the covering factor $\xi_l=1$ required by the saturation which is observed in the high resolution Keck spectrum.

Finally we stress  that, while ionisation changes can more easily explain  correlated variations in different velocity intervals of a BAL structure, in some cases  observations show different portions of a BAL varying independently  \citep{Hall11,Cape12}. The latter  are more easily interpreted as due to gas clouds crossing the line of sight. A priori there is no reason to expect that this kind of variation should show the same SF as we observe in our case, where changes in different velocity intervals are strongly correlated. In this sense, the similarity between the SF of APM 08279+5255 and the ensemble SF deduced from the literature, which likely includes cases of BAL variations caused by independent clouds, either provides statistical constraints on geometrical models, or implies that changes due to geometrical effects represent a minority of the observed cases.

\begin{acknowledgements}
It is a pleasure to acknowledge M.J. Irwin for providing us with the discovery spectrum of APM 08279+5255, D.C. Hines and G.D. Schmidt for the spectrum taken at the Steward Observatory 2.3 m Bok telescope, T.A. Barlow for sending the equivalent width variability data of his PhD thesis. We are grateful to E. Piconcelli and F. Fiore for useful discussions. We acknowledge S. Benetti and the technical staff of the Asiago Observatory for the support provided during the entire monitoring campaign.
\end{acknowledgements}

\bibliographystyle{aa}
\bibliography{trevese-v1}{}

\end{document}